\documentclass[reprint, superscriptaddress, twocolumn, amsmath, amssymb, aps, prb]{revtex4-2}
\usepackage{graphicx}
\usepackage{dcolumn}
\usepackage{bm}
\usepackage{placeins}
\usepackage{appendix}
\usepackage{upgreek}

\usepackage{verbatim}
\usepackage[usenames,dvipsnames]{xcolor}
 \usepackage{amsmath}
 \usepackage{amsfonts}
 \usepackage{amssymb}
\usepackage[colorlinks=true,citecolor=blue,linkcolor=red]{hyperref}

 \usepackage{bbold}     
 \usepackage[makeroom]{cancel}  
 \usepackage{multirow}    
 \usepackage[normalem]{ulem}        
 \usepackage{array}
 \usepackage{pdfpages}
\makeatletter
 \AtBeginDocument{\let\LS@rot\@undefined}
 \makeatother

\begin{document}


\title{Reducing disorder in PbTe nanowires for Majorana research}

\author{Wenyu Song}
\email{equal contribution}
\affiliation{State Key Laboratory of Low Dimensional Quantum Physics, Department of Physics, Tsinghua University, Beijing 100084, China}

\author{Zehao Yu}
\email{equal contribution}
\affiliation{State Key Laboratory of Low Dimensional Quantum Physics, Department of Physics, Tsinghua University, Beijing 100084, China}

\author{Yuhao Wang}
\email{equal contribution}
\affiliation{State Key Laboratory of Low Dimensional Quantum Physics, Department of Physics, Tsinghua University, Beijing 100084, China}

\author{Yichun Gao}
\email{equal contribution}
\affiliation{State Key Laboratory of Low Dimensional Quantum Physics, Department of Physics, Tsinghua University, Beijing 100084, China}

\author{Zonglin Li}
\affiliation{State Key Laboratory of Low Dimensional Quantum Physics, Department of Physics, Tsinghua University, Beijing 100084, China}

\author{Shuai Yang}
\affiliation{State Key Laboratory of Low Dimensional Quantum Physics, Department of Physics, Tsinghua University, Beijing 100084, China}

\author{Shan Zhang}
\affiliation{State Key Laboratory of Low Dimensional Quantum Physics, Department of Physics, Tsinghua University, Beijing 100084, China}

\author{Zuhan Geng}
\affiliation{State Key Laboratory of Low Dimensional Quantum Physics, Department of Physics, Tsinghua University, Beijing 100084, China}

\author{Ruidong Li}
\affiliation{State Key Laboratory of Low Dimensional Quantum Physics, Department of Physics, Tsinghua University, Beijing 100084, China}

\author{Zhaoyu Wang}
\affiliation{State Key Laboratory of Low Dimensional Quantum Physics, Department of Physics, Tsinghua University, Beijing 100084, China}

\author{Fangting Chen}
\affiliation{State Key Laboratory of Low Dimensional Quantum Physics, Department of Physics, Tsinghua University, Beijing 100084, China}

\author{Lining Yang}
\affiliation{State Key Laboratory of Low Dimensional Quantum Physics, Department of Physics, Tsinghua University, Beijing 100084, China}

\author{Wentao Miao}
\affiliation{State Key Laboratory of Low Dimensional Quantum Physics, Department of Physics, Tsinghua University, Beijing 100084, China}

\author{Jiaye Xu}
\affiliation{State Key Laboratory of Low Dimensional Quantum Physics, Department of Physics, Tsinghua University, Beijing 100084, China}

\author{Xiao Feng}
\affiliation{State Key Laboratory of Low Dimensional Quantum Physics, Department of Physics, Tsinghua University, Beijing 100084, China}
\affiliation{Beijing Academy of Quantum Information Sciences, Beijing 100193, China}
\affiliation{Frontier Science Center for Quantum Information, Beijing 100084, China}
\affiliation{Hefei National Laboratory, Hefei 230088, China}

\author{Tiantian Wang}
\affiliation{Beijing Academy of Quantum Information Sciences, Beijing 100193, China}
\affiliation{Hefei National Laboratory, Hefei 230088, China}

\author{Yunyi Zang}
\affiliation{Beijing Academy of Quantum Information Sciences, Beijing 100193, China}
\affiliation{Hefei National Laboratory, Hefei 230088, China}

\author{Lin Li}
\affiliation{Beijing Academy of Quantum Information Sciences, Beijing 100193, China}

\author{Runan Shang}
\affiliation{Beijing Academy of Quantum Information Sciences, Beijing 100193, China}
\affiliation{Hefei National Laboratory, Hefei 230088, China}

\author{Qikun Xue}
\affiliation{State Key Laboratory of Low Dimensional Quantum Physics, Department of Physics, Tsinghua University, Beijing 100084, China}
\affiliation{Beijing Academy of Quantum Information Sciences, Beijing 100193, China}
\affiliation{Frontier Science Center for Quantum Information, Beijing 100084, China}
\affiliation{Hefei National Laboratory, Hefei 230088, China}
\affiliation{Southern University of Science and Technology, Shenzhen 518055, China}

\author{Ke He}
\email{kehe@tsinghua.edu.cn}
\affiliation{State Key Laboratory of Low Dimensional Quantum Physics, Department of Physics, Tsinghua University, Beijing 100084, China}
\affiliation{Beijing Academy of Quantum Information Sciences, Beijing 100193, China}
\affiliation{Frontier Science Center for Quantum Information, Beijing 100084, China}
\affiliation{Hefei National Laboratory, Hefei 230088, China}

\author{Hao Zhang}
\email{hzquantum@mail.tsinghua.edu.cn}
\affiliation{State Key Laboratory of Low Dimensional Quantum Physics, Department of Physics, Tsinghua University, Beijing 100084, China}
\affiliation{Beijing Academy of Quantum Information Sciences, Beijing 100193, China}
\affiliation{Frontier Science Center for Quantum Information, Beijing 100084, China}


\begin{abstract}

Material challenges are the key issue in Majorana research where surface disorder constrains device performance. Here, we tackle this challenge by embedding PbTe nanowires within a lattice-constant-matched crystal. The wire edges are shaped by self-organized growth instead of lithography, resulting in nearly-atomic-flat facets along both cross-sectional and longitudinal directions. Quantized conductance is observed at zero magnetic field with channel lengths maximally reaching 1.7 $\upmu$m, significantly surpassing the state-of-the-art of III-V nanowires (an order-of-magnitude improvement compared to InSb). Coupling PbTe to a Pb film unveils a flat interface spanning microns and a large superconducting gap of 1 meV. Our result not only represents a stride toward meeting the stringent low-disorder requirement for Majoranas, but may also open the door to various hybrid quantum devices requiring a low level of disorder.

\end{abstract}

\maketitle  

Quantum computing technologies have advanced to a stage where deeper involvement of materials science is highly desired \cite{2021_Science_Review}. The microscopic mechanisms of qubit decoherence can be traced back to poor control of surfaces and interfaces in corresponding devices. Surface oxides and interface inhomogeneity can significantly degrade device quality. Reducing those disorder is a challenging task that requires new interdisciplinary approaches. Take, for instance, semiconductor nanowires (InAs and InSb), a material platform extensively studied for Majorana zero modes \cite{Lutchyn2010, Oreg2010, Mourik, Deng2016, Nichele2017, Gul2018,  Song2022, WangZhaoyu,Delft_Kitaev, MS_2023, NextSteps}. For free-standing wires, despite the bulk being of single crystal, the surface facets are covered by undesired oxide \cite{Gul2017, PanCPL}. For in-plane wires, additional sources of disorder can emerge, including lattice mismatch and interface inhomogeneity \cite{2018_PRM_SAG,Palmstrom_SAG, Pavel_SAG_2019, Roy}.  For gate-defined wires based on two dimensional electron gases (2DEGs), lithographic inhomogeneity is a severe disorder source \cite{Nichele2017, MS_2023}. Disorder effects can mimic and destroy Majorana signatures \cite{Patrick_Lee_disorder_2012, Prada2012, Loss2013ZBP, Liu2017, GoodBadUgly, DasSarma_estimate, DasSarma2021Disorder, Tudor2021Disorder}, greatly hindering experimental progress.

Selective-area-grown (SAG) PbTe nanowires have recently been proposed as a possible Majorana platform \cite{CaoZhanPbTe}. Rapid experimental progress has been reported \cite{CaoZhanPbTe, Jiangyuying, PbTe_AB,  Zitong, Erik_PbTe_SAG, Fabrizio_PbTe, Wenyu, Yichun, Yuhao, Ruidong, Frolov_PbTe}. The appealing advantage of PbTe lies in its large dielectric constant ($\sim$ 1350), which can mitigate disorder-induced potential fluctuations. Additionally, the lattice-constant-matched CdTe substrate eliminates a major disorder source for in-plane wires \cite{Jiangyuying, PbTe_AB, Zitong}. Furthermore, capping PbTe with CdTe spatially separates the device from surface oxide. These material merits, however, were not manifested in device transport, as the prior PbTe devices \cite{Jiangyuying, PbTe_AB, Zitong} did not show superior transport compared to III-V wires. The persistence of other disorder sources thus necessitates further elimination. 

In this paper, we identify interface inhomogeneity as the dominant disorder in SAG PbTe nanowires, and present our material approach to address this issue. Step by step, we tackled the disorder issue on the bottom and side facets, yielding nearly atomic-flat interfaces. Consequently, ballistic transport can be routinely achieved at zero magnetic field, evidenced by the observed conductance quantization. The channel length can remarkably reach 1.7 $\upmu$m, an-order-of-magnitude improvement compared to the state-of-the-art in InSb. For free-standing InSb wires, zero-field quantization is observed for channel lengths ranging from 80$-$170 nm \cite{Kammhuber2016, Zhang2017Ballistic, 2017_Jakob_helical}, beyond which the plateau is suppressed. For SAG InSb, zero-field quantization has yet to be observed \cite{Pavel_SAG_2019, Roy}. For InAs wires (both free-standing and SAG), zero-field quantization is absent in most of previous works except for one (maximum channel length of 900 nm) using template assisted growth \cite{InAs_Gooth}. We then couple PbTe to Pb, employing the refined recipe, and observe a hard gap. The gap size of $\sim$ 1 meV, significantly larger than that of Al \cite{Chang2015, Krogstrup2015}, can result in a shorter coherence length, holding promise toward observing Majorana correlations over a 1.7-$\upmu$m-long channel in future devices. In addition, these nanowires may serve as a low-disordered hybrid material platform for studying exotic quantum devices such as gate-tunable qubits and Cooper pair splitters \cite{2015_PRL_gatemon, 2021_Devoret_Science, Nature_Cooper_pair}.

\begin{figure}[htb]
\includegraphics[width=\columnwidth]{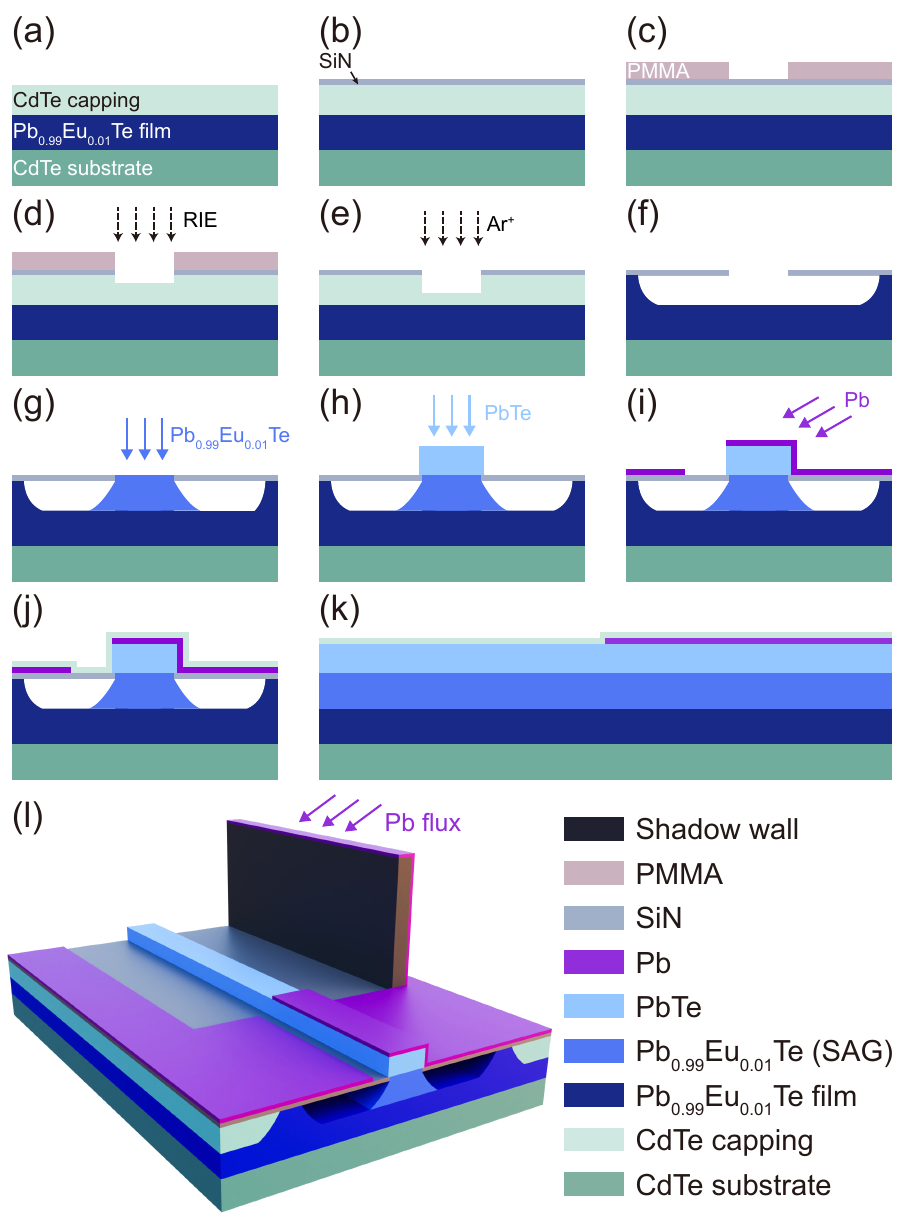}
\centering
\caption{Schematic of device growth. (a) Growth of global Pb$_{0.99}$Eu$_{0.01}$Te (middle) and CdTe capping (top) on a CdTe substrate. (b) Deposition of SiN dielectric. (c) PMMA coating with opened windows.  (d) Etching the SiN mask and part of CdTe. (e) Ar ion treatment. (f) Substrate annealing. (g) SAG Pb$_{0.99}$Eu$_{0.01}$Te buffer. (h) SAG  PbTe. (i) Pb deposition. (j) CdTe capping. (k) Device schematic along the longitudinal direction. (l) 3D schematic after step (i). }
\label{fig1}
\end{figure}

\begin{figure}[htb]
\includegraphics[width=\columnwidth]{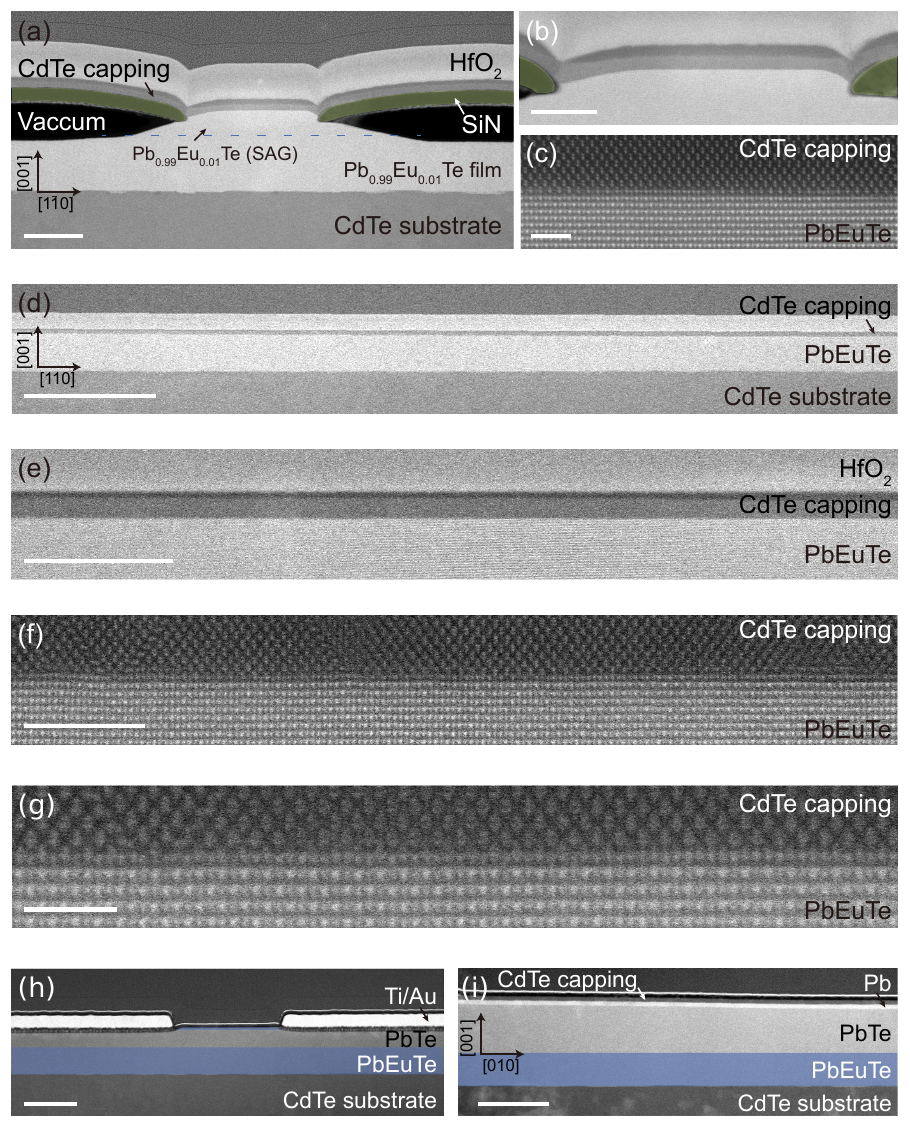}
\centering
\caption{Pb$_{0.99}$Eu$_{0.01}$Te surface. (a) Cross-sectional STEM of a test device. SiN is false-colored as green. (b-c) Enlargements of the surface. (d) STEM along the longitudinal direction. (e-g) Enlargements of (d). (h-i) STEM of two real devices. PbEuTe is false-colored blue. Scale bars in (a-i): 100, 50, 2, 500, 100, 5, 2, 200, and 200 nm. }
\label{fig2}
\end{figure}

Figure 1 shows our first-step optimization that addresses the bottom facet roughness. The devices in Ref. \cite{Wenyu, Yichun, Yuhao, Ruidong} were grown following this recipe. A CdTe(001) substrate was loaded into a molecular-beam-epitaxy (MBE) chamber. The surface then underwent Ar ion sputtering to remove oxides, followed by annealing at 260-330 $^{\circ}$C to ensure surface flatness. Pb$_{0.99}$Eu$_{0.01}$Te \cite{PRB_2005_PbTe_QPC} was epitaxially grown at 310-315 $^{\circ}$C (70-nm-thick), and capped by CdTe at room temperature (60-nm-thick), as shown in Fig. 1(a). The capping is crucial: Its absence renders Pb$_{0.99}$Eu$_{0.01}$Te conductive in subsequent processing. The chip was then covered by SiN (thickness, 25-30 nm) through plasma-enhanced chemical vapor deposition (Fig. 1(b)). Nanowire-shaped trenches were delineated by etching the SiN and part of the CdTe via electron beam lithography (EBL) and reactive ion etching (RIE), see Figs. 1(c-d).

After dissolving the electron-beam resist (PMMA), the chip was cleaned through oxygen plasma, and reloaded into the MBE chamber for Ar ion treatment to remove surface oxides (Fig. 1(e)). The substrate was then heated to 345 $^{\circ}$C to decompose the residual CdTe capping (Fig. 1(f)). Subsequently, a fresh layer of Pb$_{0.99}$Eu$_{0.01}$Te was selectively grown within the trench at 340-350 $^{\circ}$C (thickness:  20-40 nm). We have tested and confirmed the insulating nature of Pb$_{0.99}$Eu$_{0.01}$Te, whose conduction band is $\sim$ 25 meV higher than that of PbTe \cite{1997_CB_offset}. Up to Fig. 1(g), a lattice-matched flat substrate was achieved. PbTe was then grown selectively at 320-330 $^{\circ}$C, followed by Pb deposition (stage cooled by liquid nitrogen) and CdTe capping, as shown in Figs. 1(h-j). Figures 1(k-l) are the device schematics (the shadow wall was fabricated prior to the initial step). For more details, see the Supporting Information.  

Figures 1(a-g) are the key advancement compared to our old approach \cite{Jiangyuying, PbTe_AB, Zitong}, wherein PbTe was directly grown on CdTe. Since the substrate has to be heated to 320-330 $^{\circ}$C for PbTe SAG, the CdTe substrate would decompose if uncovered by PbTe, resulting in a rough PbTe-CdTe interface (see Fig. S1 in the Supporting Information). This drawback is solved in Fig. 1 by covering the CdTe substrate with PbEuTe, preventing its decomposition.

Figure 2 demonstrates the flatness of the bottom facet (Pb$_{0.99}$Eu$_{0.01}$Te). Since Pb$_{0.99}$Eu$_{0.01}$Te and PbTe are indistinguishable in scanning transmission electron microscopy (STEM), to visualize this interface, we grew ``test devices'': After Fig. 1(g), we grew CdTe instead of PbTe. Thus, the flatness of Pb$_{0.99}$Eu$_{0.01}$Te-CdTe interface in test devices, distinguishable in STEM, is an assessment of the Pb$_{0.99}$Eu$_{0.01}$Te-PbTe interface in real devices. The cross section of a test device is shown in Fig. 2(a). The vacuum voids, created in the step in Fig. 1(f), were sealed by SAG Pb$_{0.99}$Eu$_{0.01}$Te. The dashed line marks the interface between SAG Pb$_{0.99}$Eu$_{0.01}$Te and global Pb$_{0.99}$Eu$_{0.01}$Te. The dielectric layer, HfO$_2$, was sputtered after growth for subsequent STEM processing.  An enlargement of the  Pb$_{0.99}$Eu$_{0.01}$Te-CdTe interface (Fig. 2(b)) shows well-defined facets. Atomic resolution of the facet (Fig. 2(c) demonstrates its flatness. The lattice constant of Pb$_{0.99}$Eu$_{0.01}$Te (0.6463 nm) matches well with PbTe (0.6462 nm)  and CdTe (0.6480 nm),  making it an excellent insulating substrate \cite{97_PRB_PbEuTe}. For enlargements over all facet sites, see Fig. S2 in the Supporting Information.

We next performed STEM along the wire's longitudinal direction of a test device, see Fig. 2(d). The Pb$_{0.99}$Eu$_{0.01}$Te-CdTe interface is nearly atomically flat over a length scale of 5 microns, see Figs. 2(e-g) and Fig. S3 in the Supporting Information for enlargements. This long-range flatness leads to the recent transport progress \cite{Wenyu, Yichun, Yuhao, Ruidong}, and is in sharp contrast to the old recipe, wherein the fluctuations of the interface are in tens of nanometers (Fig. S1).

Figure 2(h) shows STEM of a ``real device'', cut along the longitudinal direction. This device exhibits ballistic transport \cite{Yuhao}. The PbEuTe region is highlighted in blue, with thickness estimated based on the growth time. Figure 2(i) is the STEM of a PbTe-Pb nanowire, grown under identical condition as the hard gap devices in Ref. \cite{Yichun}. The PbEuTe-PbTe interface in Figs. 2(h-i) should be nearly-atomic-flat based on the test devices. For more STEM analysis, see Fig. S4 in the Supporting Information.   

\begin{figure}[ht]
\includegraphics[width=\columnwidth]{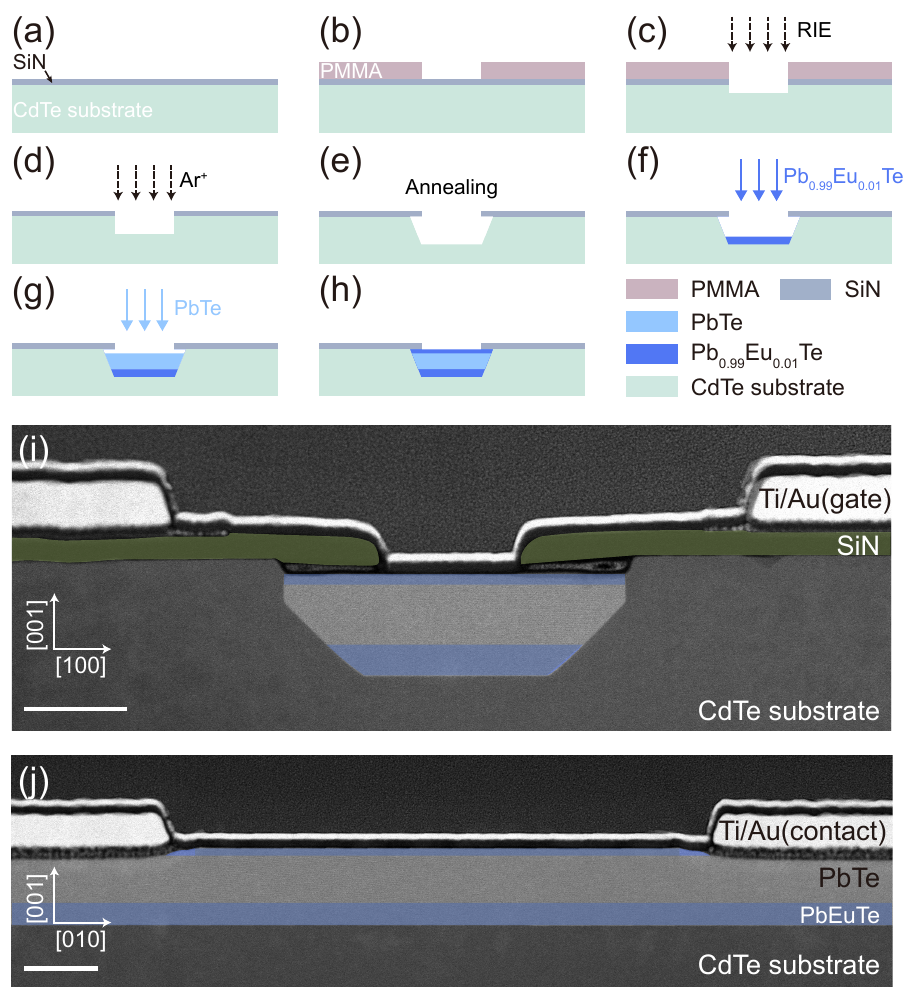}
\centering
\caption{Improved approach. (a) SiN deposition on a CdTe substrate. (b-c) Wire trenches defined via EBL and RIE. (d) Argon ion treatment. (e) Substrate annealing. (f) SAG Pb$_{0.99}$Eu$_{0.01}$Te buffer. (g) SAG PbTe. (h) SAG Pb$_{0.99}$Eu$_{0.01}$Te capping. (i) Cross section of a PbTe device. STEM of this device has been used and shown in our previous work for other purpose \cite{Yuhao_degeneracy}. (j) A PbTe device, cut along the longitudinal direction. Scale bars in (i-j), 100 nm. For (i-j), A thin layer of Ti/Au was evaporated in after measurement to facilitate the STEM preparation.   }
\label{fig3}
\end{figure}

The first step (Fig. 1) eliminates bottom facet roughness, while the roughness in side facets remains. Unlike free-standing wires, where side facets align with low-energy crystal planes and can achieve atomic flatness, the edges of SAG wires are shaped by the trenches thus inherently rough due to EBL and RIE. The trench shape affects the wire edges even without their direct contact. Another issue of Fig. 1 is the global distribution of Pb$_{0.99}$Eu$_{0.01}$Te on the substrate. Its large dielectric constant significantly increases the gate-wire capacitance coupling, resulting in the local gate behavior akin to a global gate (see Fig. S5 in the Supporting Information for transport tests). This cross-talk must be minimized in hybrid devices, where local gates should tune local potential.

Built upon Fig. 1, we then improved our approach by removing global Pb$_{0.99}$Eu$_{0.01}$Te to minimize the cross-talk, and keeping SAG Pb$_{0.99}$Eu$_{0.01}$Te to retain the bottom facet flatness, see Fig. 3 for the second-step optimization. A CdTe substrate was covered by a SiN mask with etched trenches (Figs. 3(a-c)). The chip was then loaded into an MBE chamber for oxide removal (Fig. 3(d)). The pivotal steps are Figs. 3(e-h). The substrate was heated to 345 $^{\circ}$C for 80 min (Fig. 3(e)) to uniformly decompose CdTe, forming flat facets aligned with low-energy crystal planes ($\{$001$\}$ and $\{$101$\}$). Then, a thin layer of Pb$_{0.99}$Eu$_{0.01}$Te was selectively grown at 345 $^{\circ}$C (Fig. 3(f)). The growth is uniform, as nucleation sites densely covered the bottom facet, preventing it from decomposition. Without this step, PbTe would nucleate at several discrete sites, leaving the uncovered regions of the substrate prone to decomposition, ultimately leading to substrate inhomogeneity (see Figs. S1 and S6(c)). The dramatic difference of nucleation patterns between Pb$_{0.99}$Eu$_{0.01}$Te and PbTe is probably due to the strong adhesion of Eu. Unlike III-V SAG that favors single-site nucleation to avoid grain boundary formation, the nearly matched lattice constant guarantees single-crystallinity of PbTe for multi-nucleation growth.

PbTe was then selectively grown on the flat Pb$_{0.99}$Eu$_{0.01}$Te buffer at 330-334 $^{\circ}$C (Fig. 3(g)) (Eu starts to diffuse above 400 $^{\circ}$C \cite{Eu_diffusion}). The bottom facet did not decompose due to Pb$_{0.99}$Eu$_{0.01}$Te coverage. The CdTe side facets decomposed in a uniform manner, as the PbTe grew layer by layer. The resulting side facets thus exhibit a homogeneous slope. The wire was then capped by Pb$_{0.99}$Eu$_{0.01}$Te (Fig. 3(h)). The three SAG layers in Figs. 3(f-h) should be thin, so that the device can be embedded within the substrate with a matched lattice constant, and spatially isolated from the mask, see Fig. 3(i) for a device STEM. The SiN mask is in green. Figure S7 in the Supporting Information shows atomically-resolved STEM around the side edges. The wire edges are no longer defined by lithographic trenches due to the spatial separation, but formed via self-organized growth, solving the roughness issue of side facets. Note that lithographic inhomogeneity is prevail for all SAG and 2DEG-based wires. 

Figure 3(j) is a device STEM along the wire axis. The PbTe channel between Ti/Au contacts is well-protected by Pb$_{0.99}$Eu$_{0.01}$Te. The devices in Figs. 3(i-j) are both ballistic, and were grown using the recipe in Fig. 3. Both the top and bottom facets of Pb$_{0.99}$Eu$_{0.01}$Te are flat in ``test devices'', see Fig. S6. For an overview of our step-by-step optimization and device statistics, see Fig. S8.

Figure 4 presents device transport at zero field, after the second-step optimization. Two types of devices were measured (temperature $<$ 50 mK) with their scanning electron micrographs (SEMs) displayed in Figs. 4(a-b). The first is field-effect devices (Fig. 4(a)) without superconductor. Figure 4(c) shows the conductance, $G\equiv dI/dV$, as a function of gate voltage ($V_{\text{G}}$) for 11 devices (the blue curve is the device in Fig. 3(j)). $V_{\text{G}}$'s are offset  for clarity, see Fig. S9 in the Supporting Information for the curves without offset. $V$ is the bias across the device (fixed at 0 mV in Fig. 4(c)), and $I$ is the current. The channel lengths, i.e. the spacing between contacts, range from 690 nm to 1.7 $\upmu$m. Quantized conductance plateaus at multiples of $2e^2/h$ are observed. The plateau widths varies due to variation in wire thickness (subband spacing) and geometric capacitance for different devices. The channel length of zero-field quantization reaching 1.7 $\upmu$m (Fig. 4(a)) is a hallmark indicator that the disorder level in PbTe is significantly lower than the best-optimized InAs and InSb nanowires. This length also exceeds the ballistic length (maximum $\sim$ 350 nm) in devices grown using the approach of Fig. 1 \cite{Yuhao} by a factor of 4-5, signifying and quantifying the important role of disorder mitigation on lithographic inhomogeneity. The absence of the first plateau in some devices indicates subband degeneracy \cite{Yuhao_degeneracy}. Figure 4(d) shows a 2D map of the 1.7-$\upmu$m device, where quantizations are revealed as diamond shapes. The diamond size, $\sim$ 5.5 meV, measures the subband spacing. For 2D maps of other devices and magnetic field scans, see Fig. S9.    

\begin{figure}[ht]
\includegraphics[width=\columnwidth]{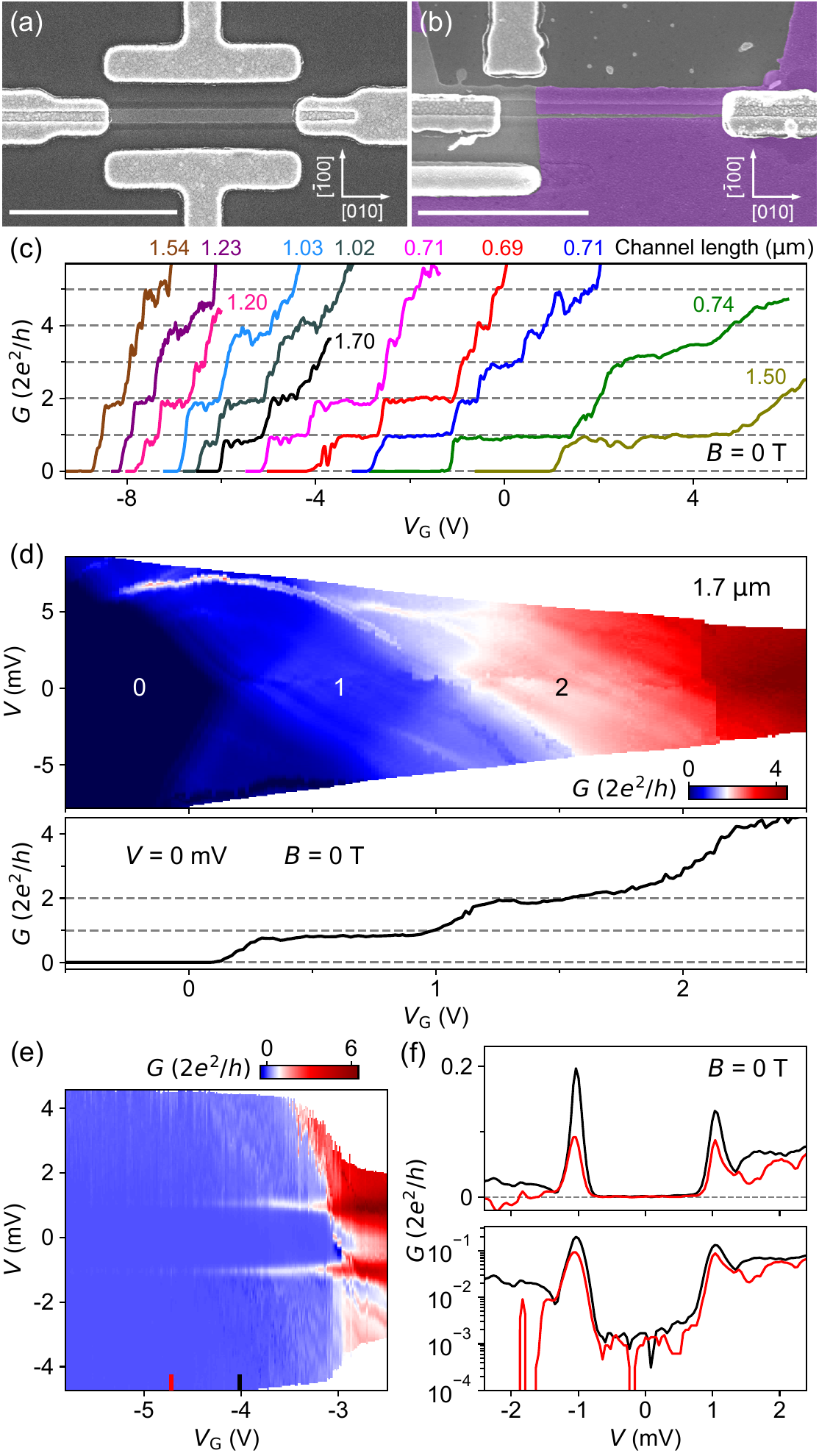}
\centering
\caption{(a) SEM of a PbTe device with a channel length of 1.7 $\upmu$m (measurement uncertainty: 50 nm).  (b) A PbTe-Pb device (Pb film false-colored violet). Scale bars in (a-b), 1.5 $\upmu$m. (c) $G$ vs $V_{\text{G}}$ of 11 devices with channel lengths labeled. (d) $G$ vs $V$ and $V_{\text{G}}$ of the device in (a). Labeled numbers represent plateau values in units of $2e^2/h$. Lower panel, zero-bias line cut. (e) $G$ vs $V$ and $V_{\text{G}}$ of the device in (b). (f) Vertical line cuts of (e), indicated by the color bars, in linear (upper) and logarithmic scales (lower). }
\label{fig4}
\end{figure}

A local constriction in the channel could, in principle, also yield quantized conductance. This scenario is highly unlikely given that the whole channel is quite uniform (nearly atomic flat). In addition, the gates span the entire channel uniformly, further preventing the formation of a local constriction. In fact, local constrictions are more likely to form in devices grown using the old recipes (Fig. 1 and Fig. S1), where zero-field quantization was not observed for long-channel devices.

Figures 4(e-f) show a PbTe-Pb device (SEM shown in Fig. 4(b)), revealing a hard gap in the tunneling regime. The subgap conductance fluctuations are equipment noise. The gap size, $\Delta \sim$ 1 meV, is five times larger than that of Al. This value is also larger than that ($\sim$ 0.42 meV) of PbTe-Pb wires grown using the old approach in Fig. S1 \cite{Zitong}. This hard gap device has minimal gate cross-talk, desired for Majorana exploration, compared to devices in Ref \cite{Yichun}. The device STEM reveals a flat PbTe-Pb interface spanning over 4 microns, see Fig. S10. 

To conclude, we report material approaches on disorder mitigation in PbTe nanowires. The inclusion of SAG PbEuTe eliminates the substrate roughness, resulting in a nearly atomic-flat bottom facet. By controlling CdTe decomposition, the PbTe wire can be embedded into the CdTe crystal with flat side facets. This spatial isolation of wire edges from the mask trenches eliminates the lithographic inhomogeneity, a major disorder source for SAG and 2DEG wires. These advancements result in ballistic PbTe nanowires at zero magnetic field with channel lengths reaching 1.7 $\upmu$m. This length scale significantly surpasses the state-of-the-art in III-V nanowires. Our result suggests that PbTe nanowires can be a better Majorana platform that may settle the decade-long debate rooted in disorder.

\textbf{Acknowledgment} We thank Leo Kouwenhoven for valuable comments and Yunhe Bai for technical assistance. This work is supported by  National Natural Science Foundation of China (92065206) and the Innovation Program for Quantum Science and Technology (2021ZD0302400). Raw data and processing codes within this paper are available at https://doi.org/10.5281/zenodo.14498901

\bibliography{mybibfile}

\newpage

\onecolumngrid

\newpage


\section*{Supplementary material}

# Supporting Information for "Reducing disorder in PbTe nanowires for Majorana research"

Wenyu Song,[1, *] Zehao Yu,[1, *] Yuhao Wang,[1, *] Yichun Gao,[1, *] Zonglin Li,[1] Shuai Yang,[1] Shan Zhang,[1] Zuhan Geng,[1] Ruidong Li,[1] Zhaoyu Wang,[1] Fangting Chen,[1] Lining Yang,[1] Wentao Miao,[1] Jiaye Xu,[1] Xiao Feng,[1, 2, 3, 4] Tiantian Wang,[2, 4] Yunyi Zang,[2, 4] Lin Li,[2] Runan Shang,[2, 4] Qikun Xue,[1, 2, 3, 4, 5] Ke He,[1, 2, 3, 4, †] and Hao Zhang[1, 2, 3, ‡]

[1] *State Key Laboratory of Low Dimensional Quantum Physics, Department of Physics, Tsinghua University, Beijing 100084, China*
[2] *Beijing Academy of Quantum Information Sciences, Beijing 100193, China*
[3] *Frontier Science Center for Quantum Information, Beijing 100084, China*
[4] *Hefei National Laboratory, Hefei 230088, China*
[5] *Southern University of Science and Technology, Shenzhen 518055, China*

## Additional Information on device growth and processing

### (1) Procedures of the approach in Fig. 1

**Shadow wall.** Hydrogen SilsesQuioxane (FOX-16, Dow Corning) resist was spun onto a CdTe (001) substrate at 2000 rpm for 60 s, and then baked at 150 °C for 90 s. EBL was performed to write the patterns of shadow walls and markers, followed by development in tetramethylammonium hydroxide aqueous solution for 165 s.

**Global $Pb_{0.99}Eu_{0.01}Te$ film.** The CdTe substrate was loaded into an ultra-high vacuum MBE chamber (Omicron Lab-10, base pressure $< 5.0 \times 10^{-10}$ mbar). To remove CdTe surface oxide, Ar bombardment was performed for 30 min at a beam energy of 2 keV and a beam current density of 1.47 μAcm$^{-2}$. Then, the substrate was annealed at 260 °C for 30 min, 300 °C for 10 min, and 330 °C for 10 min, respectively. A 70-nm-thick $Pb_{0.99}Eu_{0.01}Te$ film was grown at 310-315 °C using a PbTe compound source and an Eu elemental source. The ratio of Eu was estimated based on the beam flux. A 65-nm-thick CdTe layer was grown to cap the chip at room temperature.

**Dielectric trenches.** SiN dielectric (thickness: 25-30 nm) was deposited on the substrate using plasma-enhanced chemical vapor deposition. Subsequently, AR-P 672.045 resist was spun at 4000 rpm for 60 s and baked at 130 °C for 10 min. EBL was then performed to define nanowire trenches and markers. The chip was then developed in an Isopropanol (IPA) solution, MIBK: IPA=1:3, for 40 s. RIE was used to etch away the exposed SiN. The resist was dissolved in acetone, and the substrate was cleaned in oxygen plasma for 4 min.

**SAG PbTe** The pre-patterned substrate was loaded into the MBE chamber. Ar bombardment was performed for 16 min to remove CdTe surface oxide at a beam energy of 2 keV and a beam current density of 1.47 μAcm$^{-2}$ . The chip was then heated to 345 °C to decompose the remaining CdTe capping. Then, $Pb_{0.99}Eu_{0.01}Te$ was selectively grown at 340-350 °C (thickness: 20-40 nm), followed by the growth of PbTe nanowires at 320-330 °C (thickness: 40-70 nm). For field-effect devices, a 10-nm-thick $Pb_{0.99}Eu_{0.01}Te$ capping was grown. For superconducting devices, see the next step.

**Pb deposition.** The chip was in-situ transferred to another sample stage within the same MBE chamber, and was cooled to $\sim 100$ K using liquid nitrogen. A 12-nm-thick Pb film was deposited at a tilted angle (with shadowing). Finally, the entire chip was capped by a 10-nm-thick CdTe layer to prevent oxidation.

### (2) Procedures of the approach in Fig. 3

Shadow walls were fabricated on a CdTe substrate, followed by the deposition of a SiN dielectric layer. Nanowire trenches were then defined through EBL and RIE. The substrate was then loaded into the MBE chamber. Ar bombardment was performed for 18 min to remove CdTe oxide at a beam energy of 2 keV and a beam current density of 1.47 μAcm$^{-2}$. The chip was then annealed at 345 °C for 80 min. $Pb_{0.99}Eu_{0.01}Te$ was selectively grown at 345 °C (thickness: 20-40 nm). PbTe nanowires were grown at 330-334 °C (thickness: 50-110 nm). For field-effect devices, a 10-nm-thick $Pb_{0.99}Eu_{0.01}Te$ capping was grown. For superconducting devices, Pb was deposited in situ, followed by a 10-nm-thick CdTe capping.

### (3) Focused ion beam (FIB) lamellae preparation for STEM

FIB lamellae preparation may be problematic especially considering the presence of "vacuum cavities". To obtain high quality lamellae, we deposited a thin layer (60 nm) of HfO$_2$ on test devices (Fig. 2) or Ti/Au on real devices (Fig. 3) prior to the FIB process. The former was to preserve the device geometry while the later was to ensure conductivity of the substrate.

---

* equal contribution
† kehe@tsinghua.edu.cn
‡ hzquantum@mail.tsinghua.edu.cn

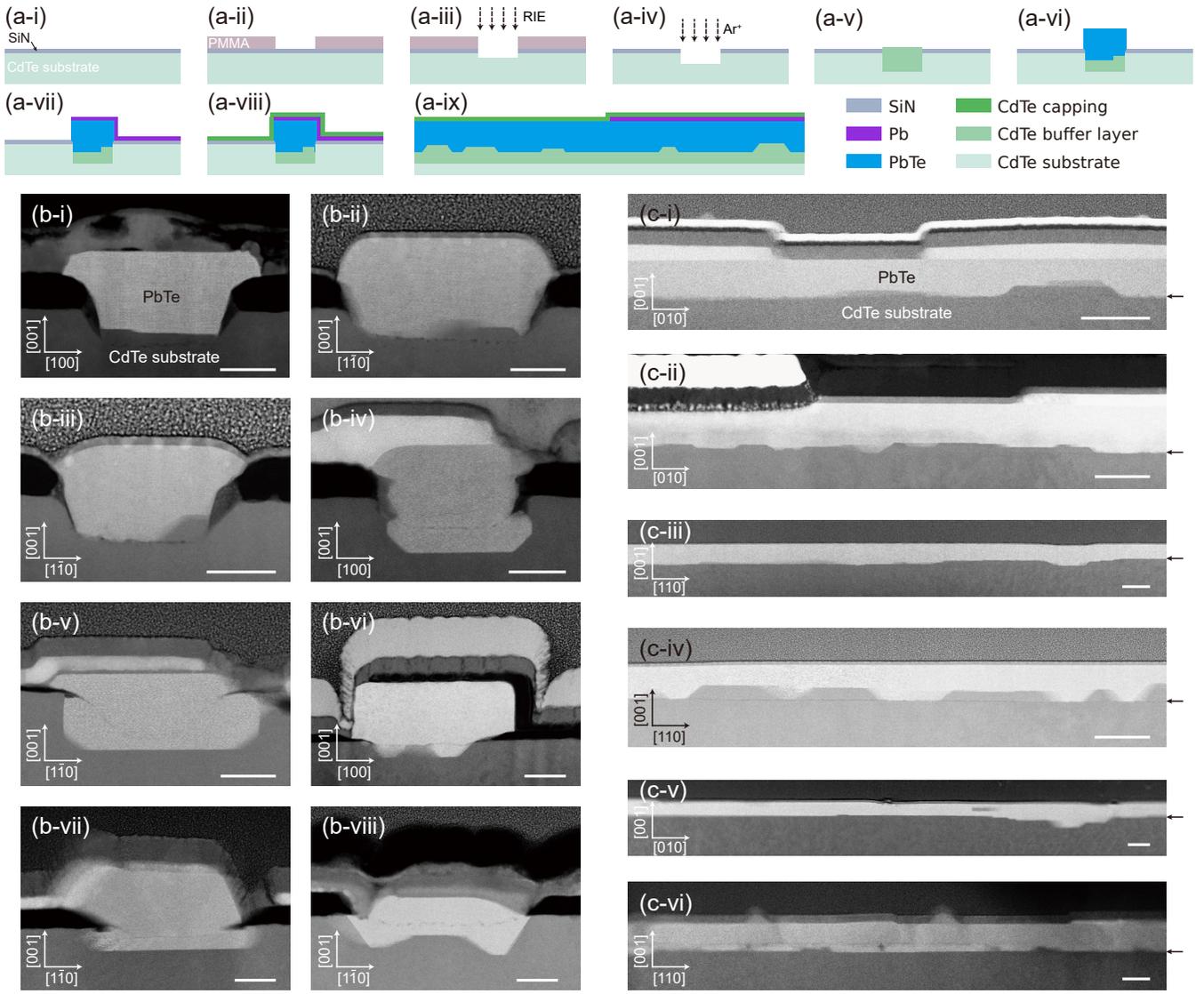

FIG. S1. (a) Growth schematic of the old approach. (a-i) Deposition of SiN on a CdTe substrate. (a-ii) PMMA-defined nanowire patterns using EBL. (a-iii) RIE to open the trenches. (a-iv) Argon ion treatment in the MBE chamber to remove CdTe oxide. (a-v) SAG CdTe buffer at 270-280 °C. (a-vi) SAG PbTe at 320-330 °C. (a-vii) Pb deposition. (a-viii) CdTe capping. (a-ix) Wire schematic along the longitudinal direction. The rough interface between PbTe and the CdTe substrate was the major disorder issue that constrained the device performance in previous works. (b) Cross-section images of devices grown using the old approach. Scale bars, 50 nm. The step of (a-v) was omitted during the growth of the devices in (b-iv)-(b-viii) . (c) High-angle annular dark field (HAADF) images of devices along the longitudinal direction. Scale bars, 100 nm. The step of (a-v) was omitted during the growth of the devices in (c-iv)-(c-vi). The arrows mark the rough PbTe-CdTe interfaces.

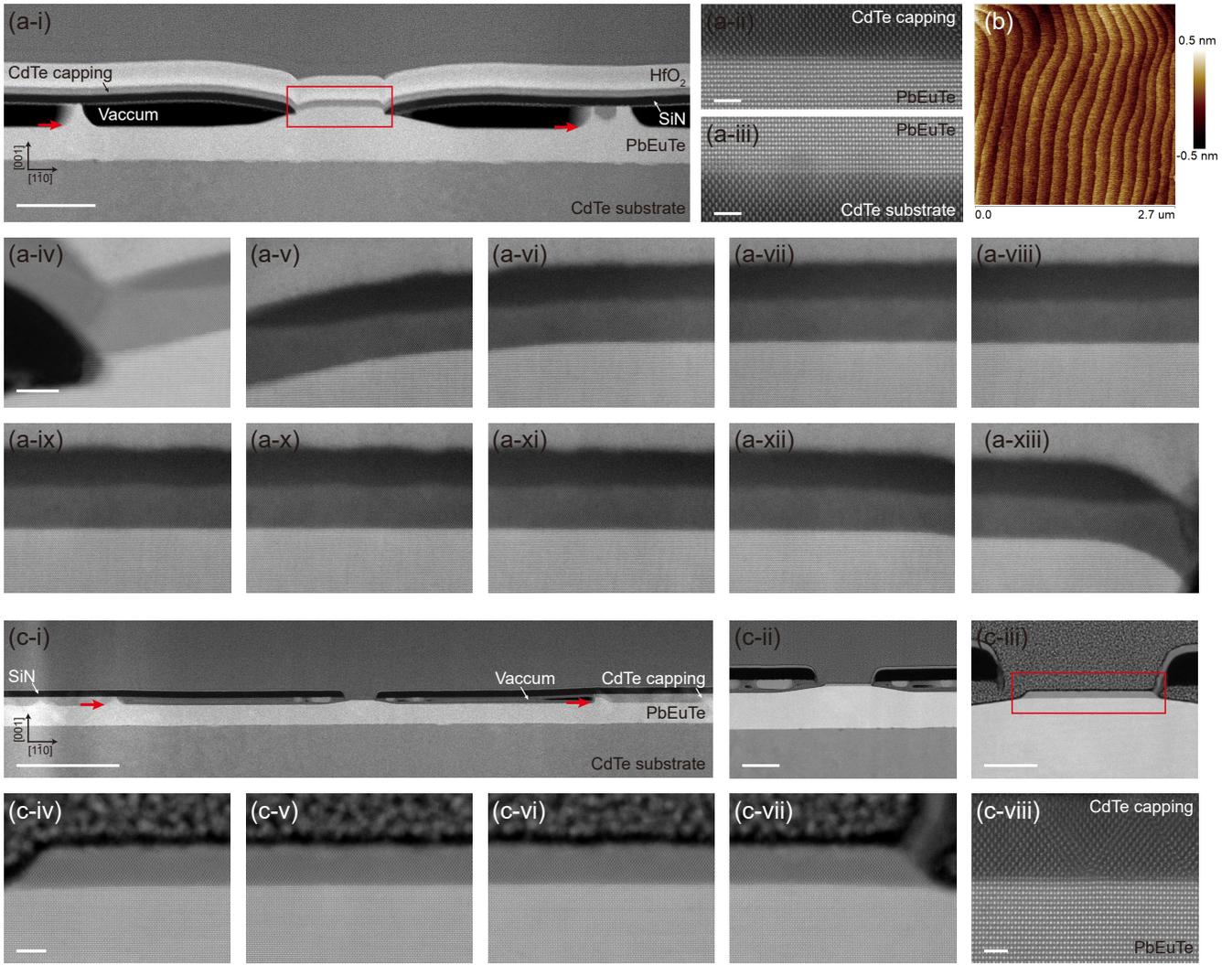

FIG. S2. (a) Additional STEM results of the test device in Figs. 2(a-c). (a-i) Cross-sectional image over a larger scale. Scale bar, 200 nm. The red arrows point to the global $Pb_{0.99}Eu_{0.01}Te$ steps, formed during the process in Fig.1(f). (a-ii) Atomically-resolved image of the $Pb_{0.99}Eu_{0.01}Te$ top surface. Scale bar, 2 nm. (a-iii) Atomically-resolved image of the global $Pb_{0.99}Eu_{0.01}Te$ bottom interface. Scale bar, 2 nm. (a-iv)-(a-xiii) Atomically-resolved images of the $Pb_{0.99}Eu_{0.01}Te$ top surface, covering all the regions indicated by the red box in (a-i). The same scale bar (10 nm) is used, shown in (a-iv). (b) Atomic force microscope image of a $Pb_{0.99}Eu_{0.01}Te$ film, grown on a CdTe(001) substrate. The film exhibits a flat, step-terraced surface. (c) STEM results of a second test device. (c-i) Cross-sectional image. Scale bar, 500 nm. (c-ii)-(c-iii), Enlargements of the $Pb_{0.99}Eu_{0.01}Te$ surface. Scale bars, 100 nm, 50 nm, respectively. Unlike the test device in panel (a), no $HfO_2$ layer was sputtered on the top of this device. Without this protection, the SiN mask was lifted during the process of lamella preparation using focused ion beam (FIB), see the left side of the red box in (c-iii). (c-iv)-(c-vii) Atomically-resolved images of the $Pb_{0.99}Eu_{0.01}Te$ surface, covering the regions indicated by the red box in (c-iii). The same scale bar (5 nm) is used, shown in (c-iv). (c-viii) Atomically resolved image of the interface between $Pb_{0.99}Eu_{0.01}Te$ and CdTe substrate. Scale bar, 2 nm.

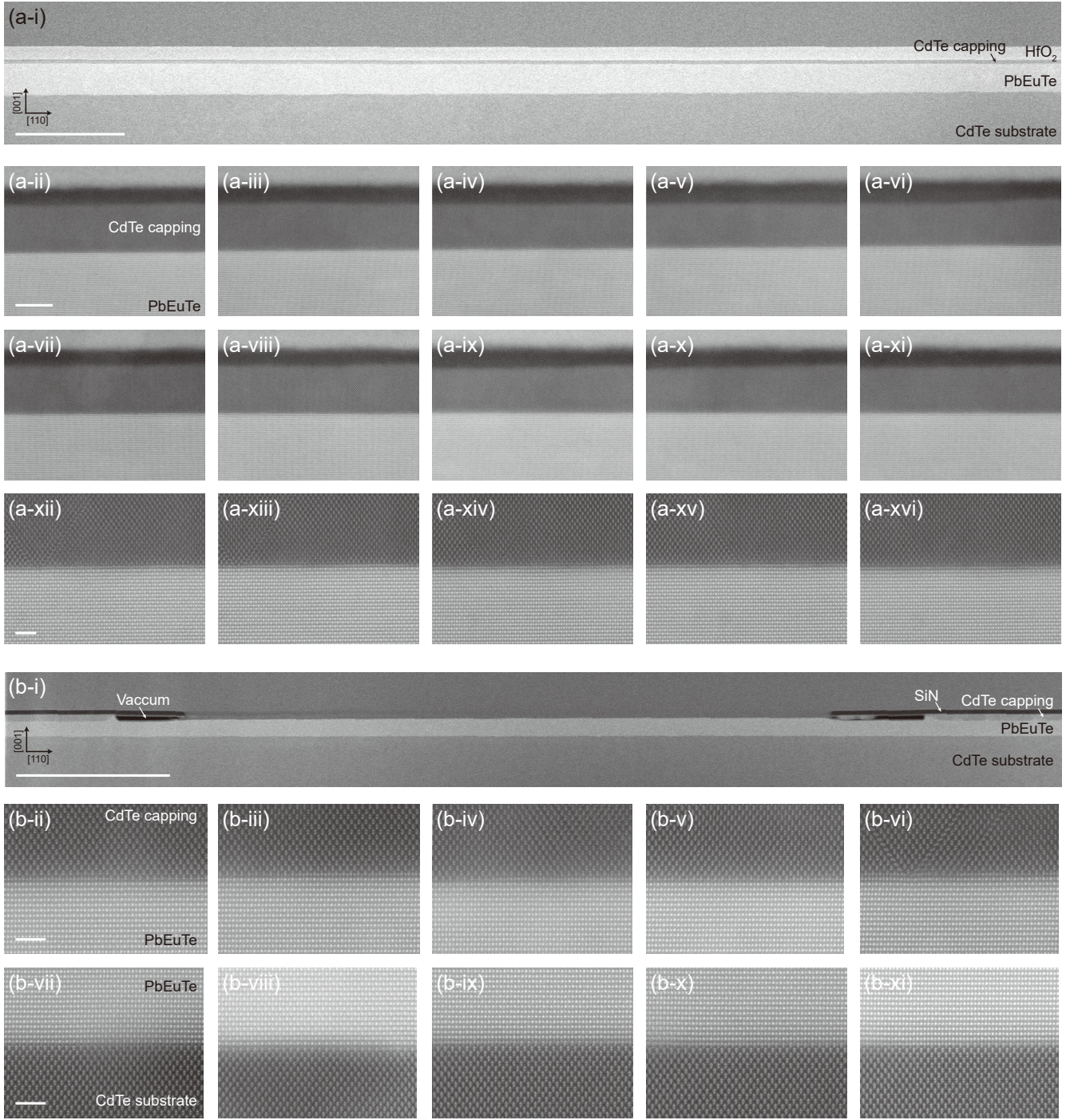

FIG. S3. (a) Additional STEM results of the test device in Figs. 2(d-g). (a-i) Image along the longitudinal direction over a larger range. Scale bar, 500 nm. (a-ii)-(a-xi) Atomically-resolved images of the $Pb_{0.99}Eu_{0.01}Te$ surface at multiple regions along the nanowire. The same scale bar (10 nm) is used, shown in (a-ii). (a-xii)-(a-xvi) Enlarged atomic resolution images of the $Pb_{0.99}Eu_{0.01}Te$ surface. The same scale bar (2 nm) is used, shown in (a-xii). (b) STEM results of another test device, cut along the longitudinal direction. (b-i) HAADF image of the test device. Scale bar, 1 μm. (b-ii)-(b-vi) Atomically-resolved images of the $Pb_{0.99}Eu_{0.01}Te$ surface at multiple regions along the nanowire. The same scale bar (2 nm) is used, shown in (b-ii). (b-vii)-(b-xi) Atomically resolved image of the interface between $Pb_{0.99}Eu_{0.01}Te$ and the CdTe substrate at multiple regions along the wire. The same scale bar (2 nm) is used, shown in (b-vii).

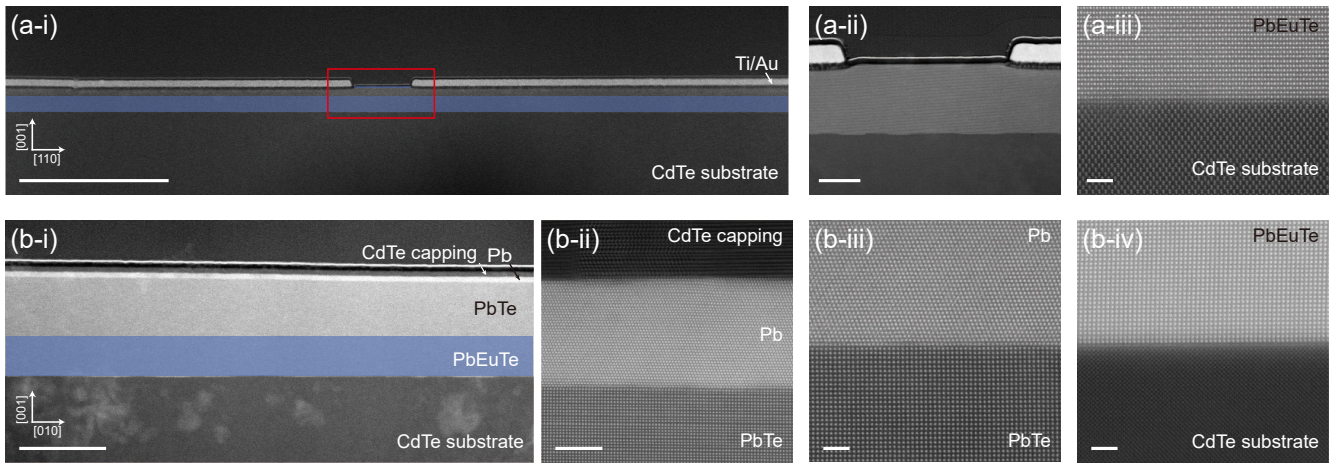

FIG. S4. (a) Additional STEM results of the device in Fig. 2(h). (a-i) Fig. 2(h) over a larger range. Scale bar, 1 μm. The $Pb_{0.99}Eu_{0.01}Te$ regions are false-colored as blue. (a-ii) Enlargement of the nanowire region between the two contacts. Scale bar, 100 nm. (a-iii) Atomically-resolved image of the interface between $Pb_{0.99}Eu_{0.01}Te$ and the CdTe substrate. Scale bar, 2 nm. (b) Additional STEM results of the device in Fig. 2(i). (b-i) Fig. 2(i) over a larger range. Scale bar, 200 nm. The $Pb_{0.99}Eu_{0.01}Te$ regions are highlighted as blue. (b-ii)-(b-iii) Atomically-resolved images of interfaces between CdTe capping, Pb, and PbTe. (b-iv) Interface between $Pb_{0.99}Eu_{0.01}Te$ and the CdTe substrate. Scale bars in (b-ii)-(b-iv), 5 nm, 2 nm, 2 nm, respectively.

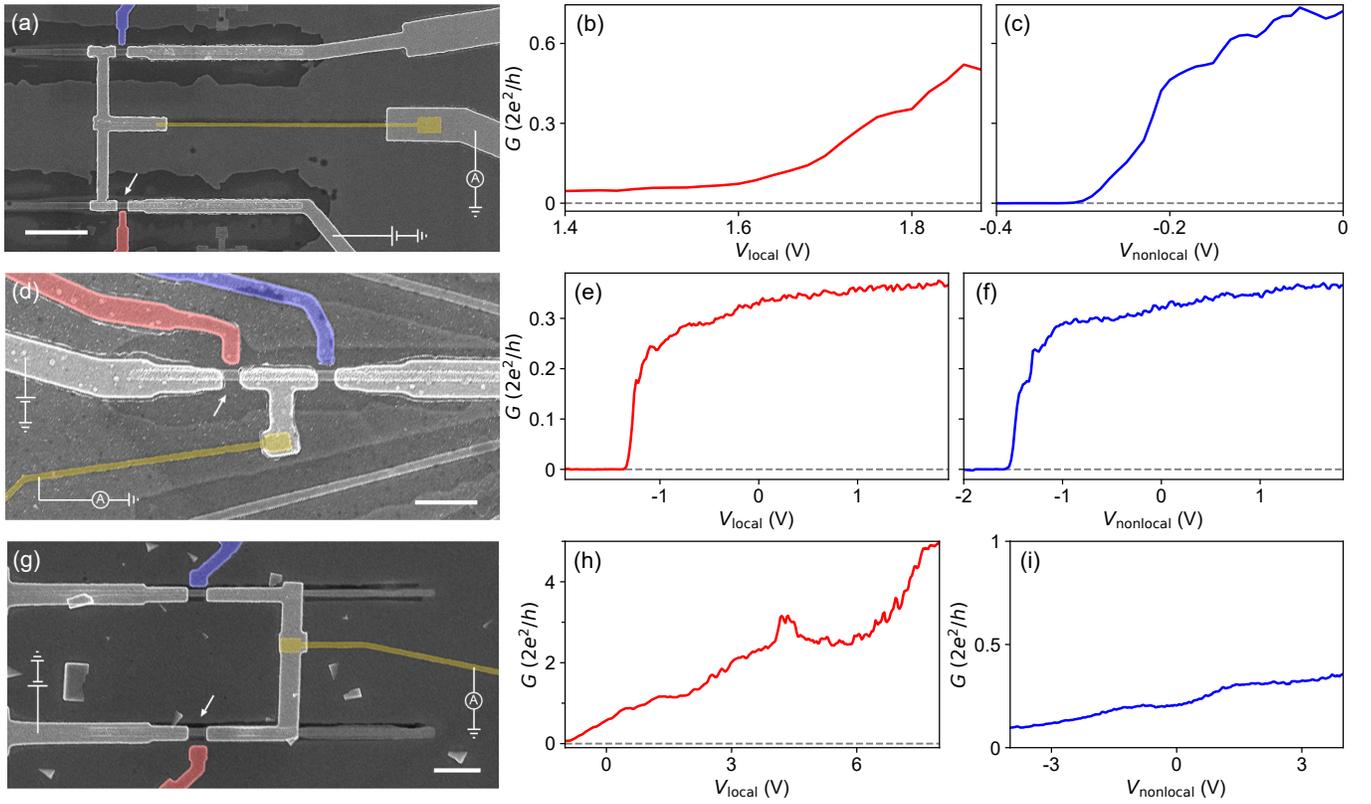

FIG. S5. Gate cross-talk test of the two approaches. (a) SEM of two parallel PbTe nanowires grown using the approach in Fig. 1. Scale bar, 2 μm. The lower wire (white arrow) was measured, while the upper wire (whose gate is false-colored blue) was kept floated. The lower two leads are the source/drain contacts (the third lead was floated). The voltage source and current meter are sketched. The orange film is Cr/Au (10 nm/2 nm). The local gate is false-colored in red (the nonlocal gate is in blue). (b-c) Conductance $G$ of the PbTe nanowire (white arrow in (a)) as a function of local gate voltage ($V_{local}$) and nonlocal gate voltage ($V_{nonlocal}$), respectively. Although the nonlocal gate is far away from the PbTe nanowire, it shows similar tunability on $G$ compared to the local gate. (d-f) Data of a second device, grown using the approach in Fig. 1, shows similar behavior: The tunability of local and nonlocal gates is almost identical. Scale bar, 1 μm. Panels (a-f) suggest that a local gate behaves like a global gate in devices grown using the approach of Fig. 1. (g-i) Data of a device grown using the approach in Fig. 3. Scale bar, 1 μm. The local gate can tune $G$ effectively, while the tunability of the nonlocal gate is 10 times smaller within the same gate voltage range (note the range difference in $y$-axis between the two panels). This ratio is roughly consistent with the ratio of geometric capacitance between the local and nonlocal gates.

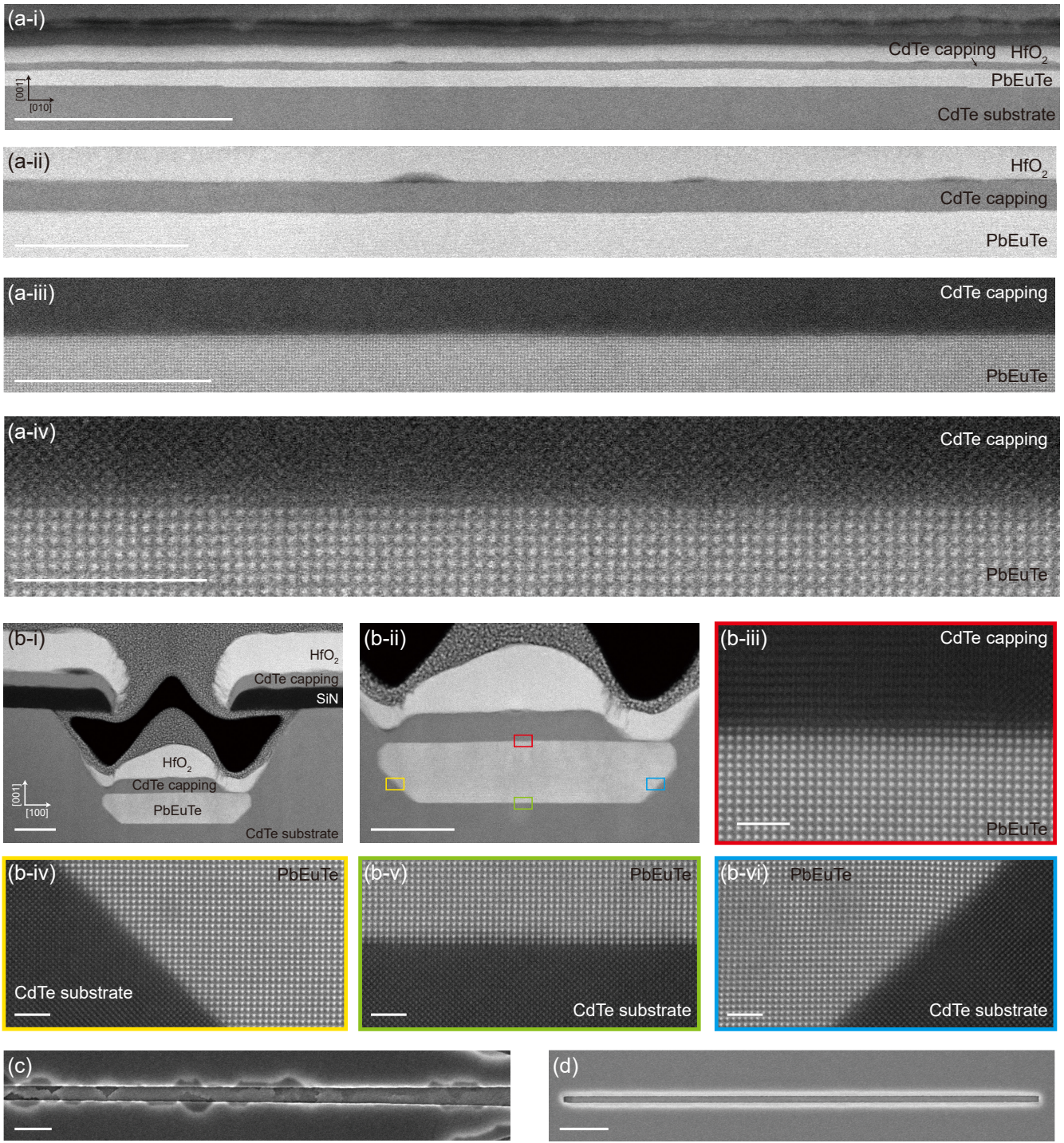

FIG. S6. (a) STEM of a "test device" of Fig. 3, cut along the longitudinal direction. CdTe capping was grown after the SAG of $Pb_{0.99}Eu_{0.01}Te$ (Fig. 3(f)). $HfO_2$ was sputtered for STEM processing. The $Pb_{0.99}Eu_{0.01}Te-CdTe$(capping) interface is flat in nanometer scale, and should be identical to the $Pb_{0.99}Eu_{0.01}Te$-PbTe interface of actual devices in Fig. 3. (a-i)-(a-iv) show the progressively enlargement of this interface. Scale bars, 500 nm, 100 nm, 20 nm, 10 nm, respectively. (b) Cross-sectional STEM of another test device. The SiN mask is highlighted in green. (b-i)-(b-ii) Cross-section images. Scale bars, 50 nm. (b-iii)-(b-vi) Atomically-resolved images of the interfaces (see the colored boxes in (b-ii)). Scale bars, 2 nm. All facets are nearly atomically flat. (c) SEM of a thin PbTe nanowire (45-nm-thick) grown using the old approach (Fig. S1). Scale bar, 400 nm. Multiple discrete nucleation sites of PbTe are visible, leading to the inhomogeneous interface. (d) SEM of a test device grown using the approach in Fig. 3: SAG $Pb_{0.99}Eu_{0.01}Te$ (30-nm-thick) on CdTe substrate. Scale bar 1 μm. A uniform flat surface is achieved due to the strong adhesion of Eu.

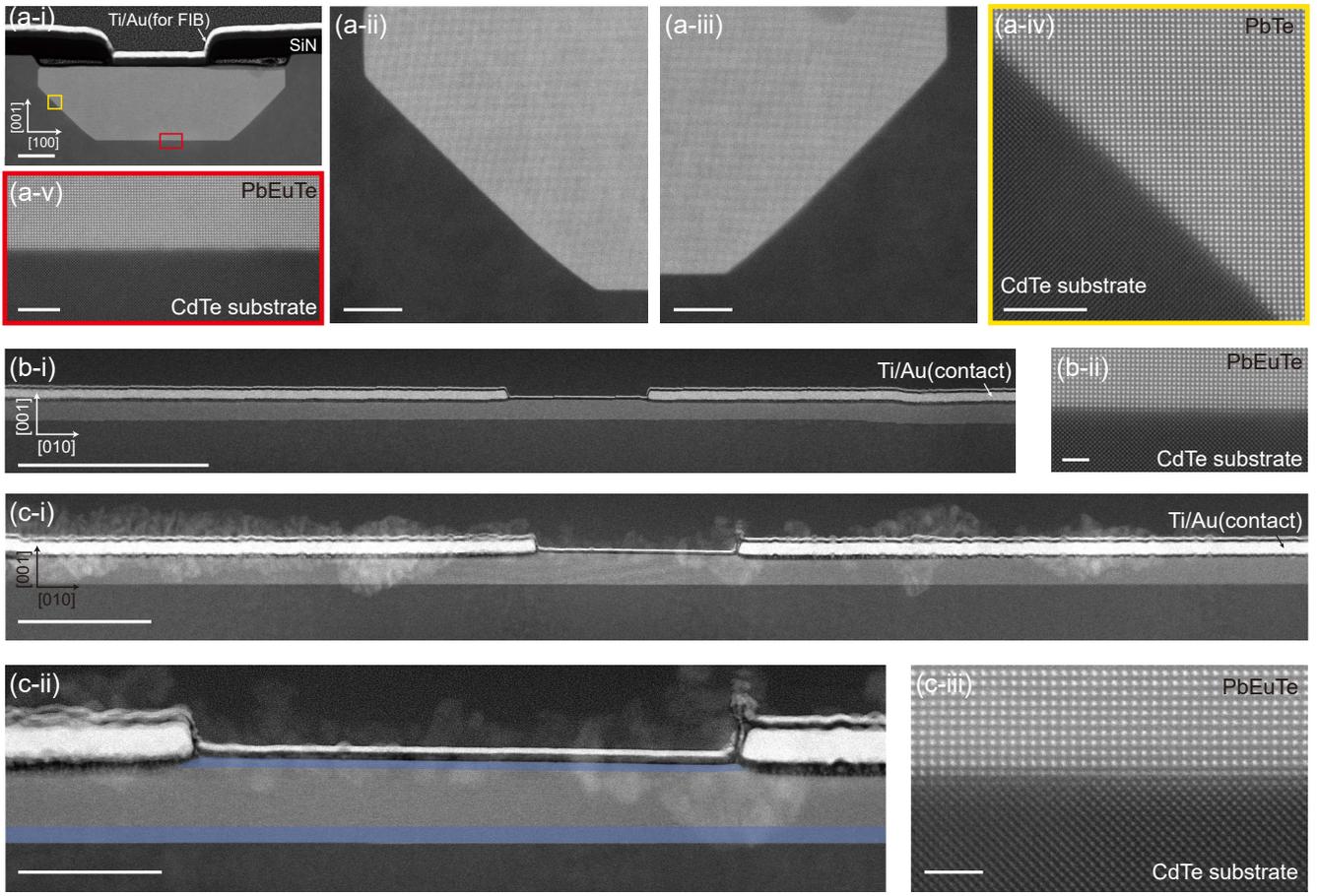

FIG. S7. (a) Additional STEM results of the device in Fig. 3(i). A thin layer of Ti/Au was deposited (see the white arrow in (a-i)) after measurement to facilitate the process of FIB. The same procedure was implemented for (b-c). (a-i) Magnified cross-section. Scale bar, 50 nm. (a-ii)-(a-iii) Side facets. Scale bars, 20 nm. (a-iv)-(a-v) Atomically-resolved images of the interfaces (see the colored boxes in (a-i)). Scale bars, 5 nm. (b) Additional STEM results of the device in Fig. 3(j). (b-i) Image over an extended range. Scale bar, 1 μm. (b-ii) Interface of the $Pb_{0.99}Eu_{0.01}Te$ and CdTe substrate. Scale bar, 2 nm. (c-i)-(c-ii) STEM of a ballistic device, cut along the longitudinal direction. Scale bars, 500 nm and 200 nm, respectively. The white patches are background contamination, created during the process of FIB. (c-iii) Atomically resolved image of the interface between $Pb_{0.99}Eu_{0.01}Te$ and CdTe substrate. Scale bar, 2 nm.

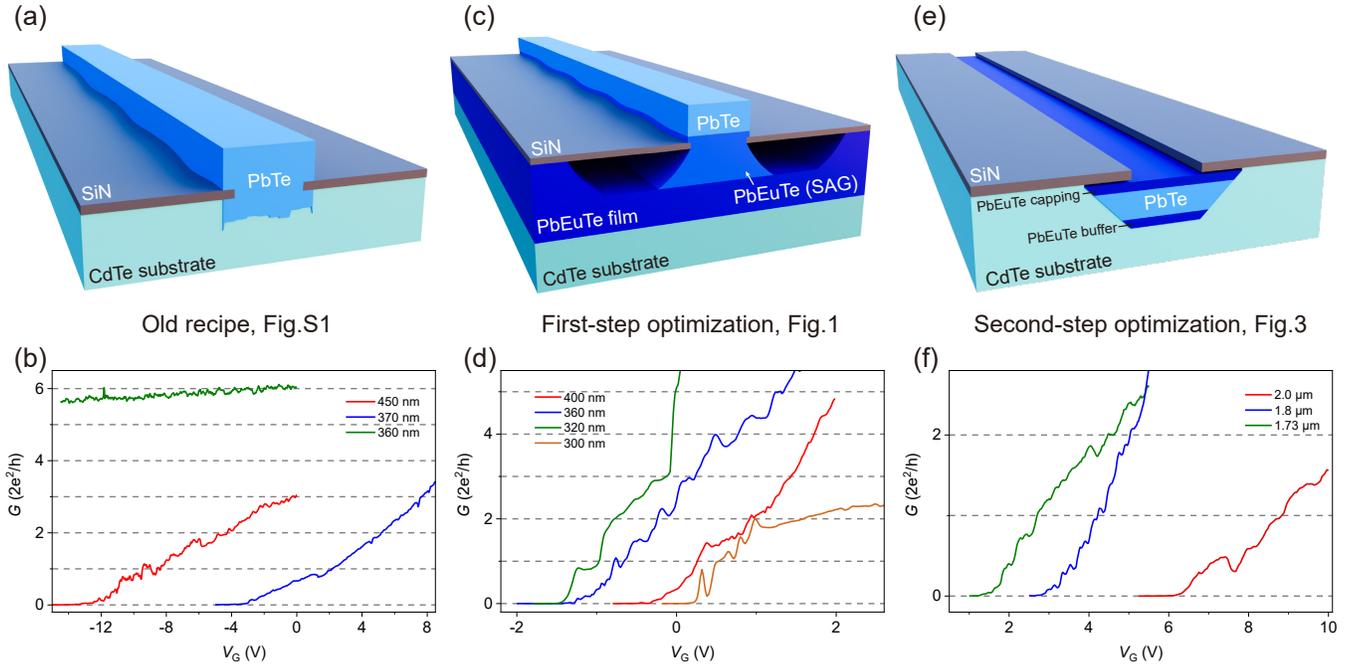

FIG. S8. Summary of the optimization effort and the corresponding device statistics. (a) Schematic of a PbTe nanowire grown using the old recipe (Fig. S1). Both the bottom facet and side facets are rough. 115 devices (short channel) were grown, fabricated, and measured under this recipe. None of them exhibited ballistic plateaus at zero magnetic field, see panel (b) for 3 examples. Many devices cannot be pinched off (see the green curve), indicating a high level of disorder. (c) Schematic of a PbTe nanowire grown after the first-step optimization (Fig. 1). The bottom facet is nearly atomically flat while the side facets are still rough due to lithography. Note that even if PbTe is not in direct contact with the SiN trench, the trench shape can still be transferred to PbTe edges via the SAG PbEuTe. 28 working devices (on a CdTe(001) substrate) were measured, while 2 of them showed zero-field quantized plateaus, with channel lengths of 320 nm (the green curve in (d)) and 350 nm. The rest 26 devices (channel lengths: 240-450 nm) did not exhibit quantized conductance (see the other three curves in (d)). Note that the device statistics provided in our previous work (Ref. [35]) include devices on both CdTe(001) substrate and CdTe(110) substrate, while the statistics here are only for the CdTe(001) case. (e) Schematic of a PbTe nanowire grown after the second-step optimization (Fig. 3). Both the bottom facet and side facets are nearly atomically flat. During the optimization, many rounds of growth, fabrication, and measurement were not successful due to various reasons. In 5 successful rounds of cool downs, 30 long-channel devices with channel lengths ranging from 0.57 μm to 1.7 μm were measured (the devices failed to conduct are not included in this statistics). 15 (out of 30) devices exhibit zero-field quantized plateaus (most of them shown in Fig. 4). Besides these 30 devices, 9 longer-channel devices (channel length: 1.73 μm to 2 μm) were also measured in those 5 cool downs. None of the 9 devices exhibit conductance quantization. Panel (f) shows 3 longer-channel devices with no quantization.

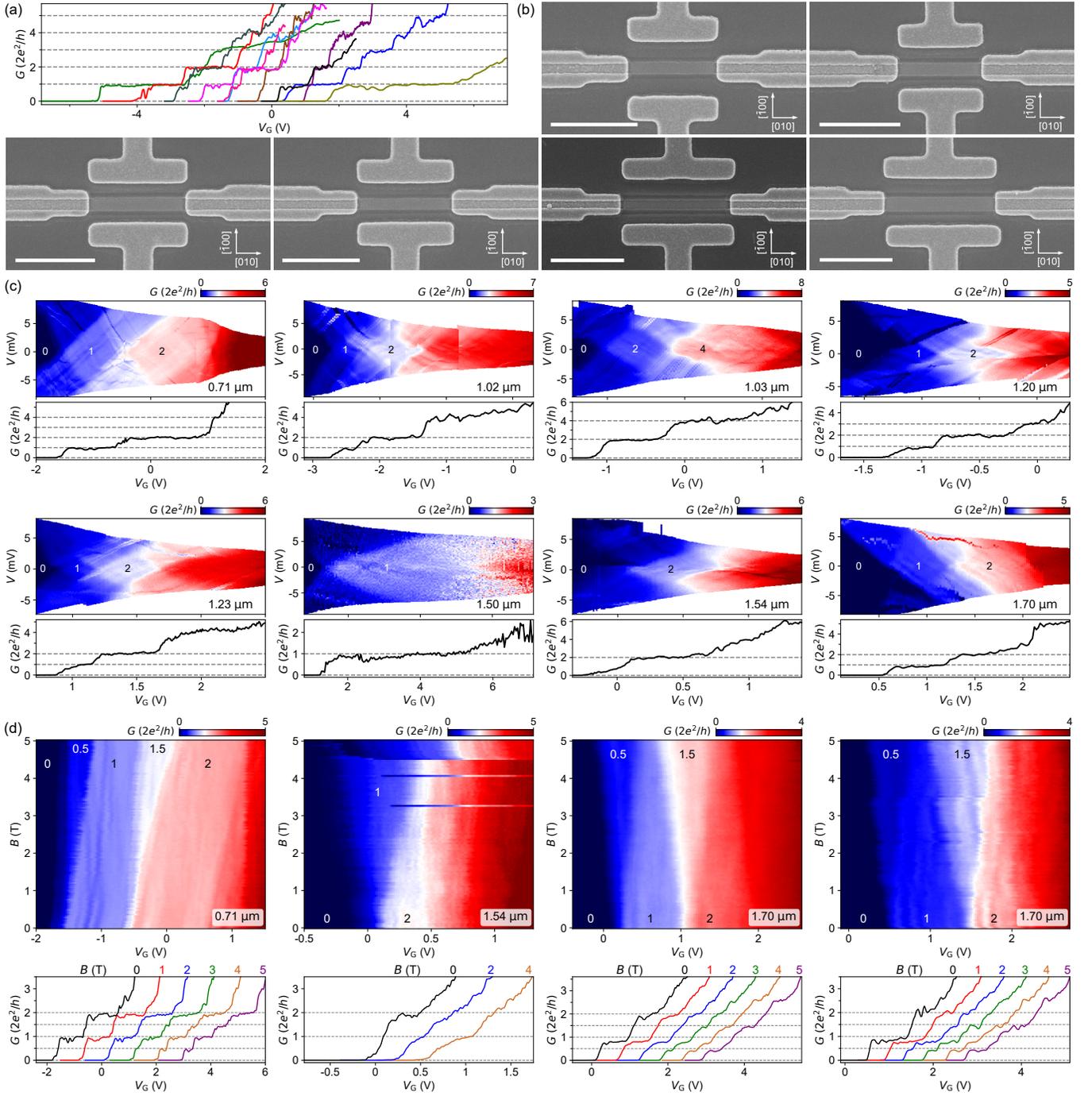

FIG. S9. (a) Replotting Fig. 4(c) without offsets in $V_G$. (b) Six additional SEMs of those devices in Fig. 4(c). Scale bars, 1 μm. Device fabrication followed the standard EBL. Before the deposition of Ti/Au contacts, PbEuTe capping in the contact regions was etched using Ar plasma for ohmic contacts. (c) $G$ vs $V$ and $V_G$ 2D maps at zero field for 8 devices (channel lengths labeled). Lower panels, zero-bias line cuts. The labeled numbers indicate the plateau diamonds (in units of $2e^2/h$). The two gates were swept simultaneously. $V_G$ is the average voltage applied to the gates. The voltage difference ($V_E$) between the gates was fixed to a value that was optimized for different devices. $V_E = $ -8 V, -6 V, and -9 V for the data in Fig. 4(c) (black curve), Fig. 4(d), and panel (c) (lower right), respectively (all corresponding to the same device). (d) $G$ vs $B$ and $V_G$ for 3 devices. The right two panels correspond to the same device (1.7 μm) with $V_E = $ -6 V and -9 V, respectively. $V = 0$ mV. $B$ is out-of-plane. Lower panels, line cuts with horizontal offsets. The subtracted contact resistances (in units of kΩ) for the 11 devices (channel lengths from 0.69 to 1.7 μm) are: 1.15, 1.15 1.05, 0 (4-terminal set-up for this particular device, 2-terminal for the rest devices), 1.4, 1.05, 1.7, 1.8, 3.0, 2.25, and 1.35. 10 out of 11 devices have dual gates.

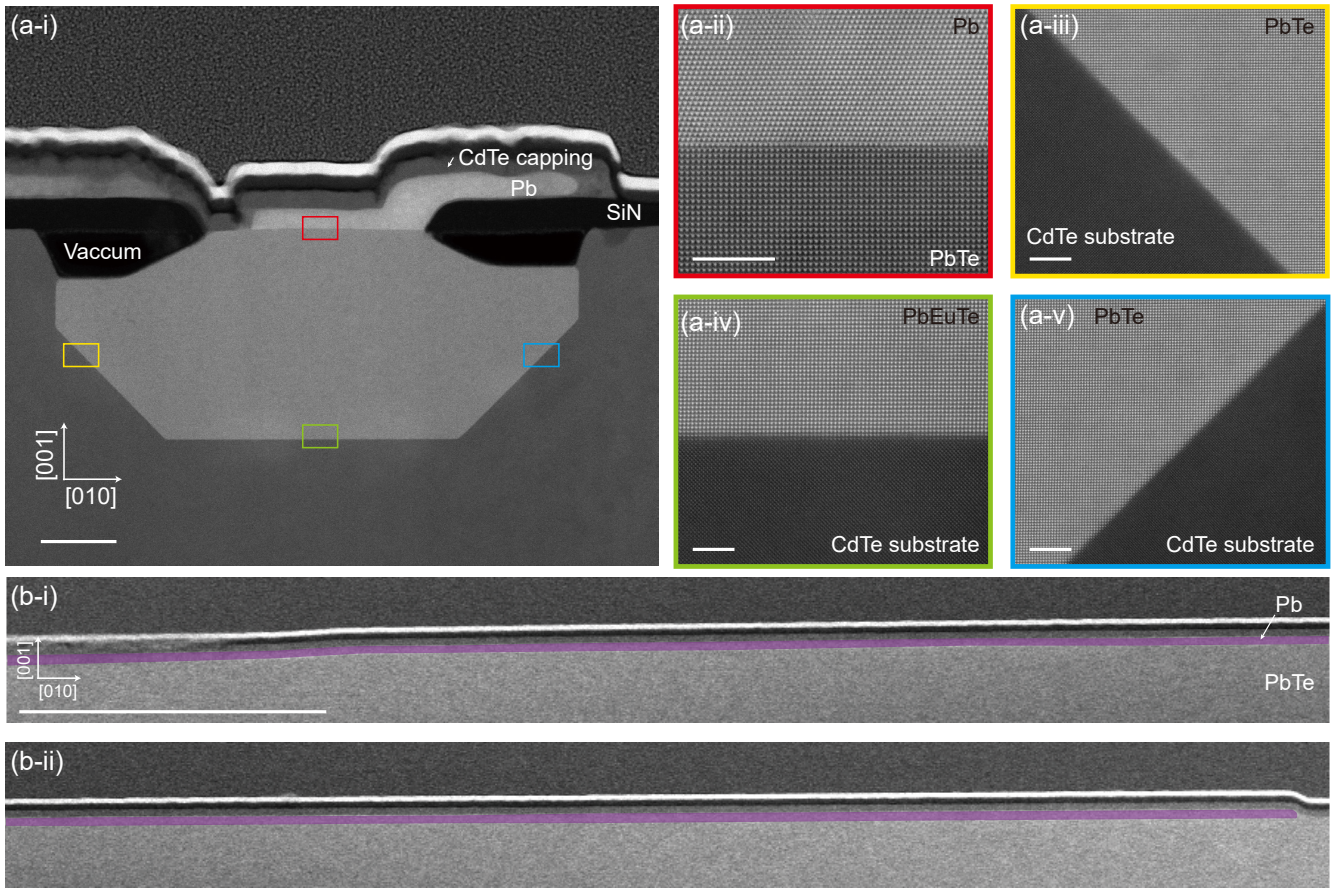

FIG. S10. (a) STEM of a hard gap device (Fig. 4(b)). (a-i) Cross-sectional image. Scale bar, 50 nm. (a-ii)-(a-v) Atomically-resolved images of interfaces (see the colored boxes in (a-i)). Scale bars, 5 nm. (b) STEM of another device (grown on the same chip as (a)), cut along the longitudinal direction. Scale bar, 500 nm. The device image is separated into (b-i) and (b-ii) for better visibility. The Pb film is flat over 4 μm. The Pb film on the substrate was etched prior to the contact fabrication to prevent a short circuit. The CdTe capping in the contact regions was removed using Ar plasma etching during device fabrication for ohmic contacts.


\end{document}